\documentclass[floatfix,%
prb,
superscriptaddress,
reprint,
preprint,
amsmath,
twocolumn,
amssymb,
amsfonts,
aps,10pt]{revtex4-2}

\usepackage{bm}% bold math
\usepackage[utf8]{inputenc}
\usepackage{etoolbox} % Provides tools for patching commands
\usepackage{enumerate}
\usepackage{enumitem}
\usepackage{graphicx}
\usepackage{mwe}
\usepackage{subfigure}
\usepackage{xcolor} %added by BF remove later
\usepackage{arydshln}
\usepackage{placeins}
\usepackage[nottoc]{tocbibind}
\usepackage{amssymb}
\usepackage{amsmath}
\usepackage{soul}
\usepackage[mathlines]{lineno}% Enable numbering of text and display math
\usepackage{appendix}
\usepackage[nottoc]{tocbibind}
\usepackage{xr}

\begin{document}

\title{Electronic Energy Scales of Cr$X_3$ ($X$ = Cl, Br, and I) using High-resolution X-ray Scattering}

\author{Chamini S. Pathiraja}
\affiliation{Physics Department and Texas Center for Superconductivity, University of Houston, Houston, TX 77204}

\author{Jayajeewana N. Ranhili}
\affiliation{Physics Department and Texas Center for Superconductivity, University of Houston, Houston, TX 77204}

\author{Deniz Wong}
\author{Christian Schulz}
\affiliation{Helmholtz-Zentrum Berlin für Materialien und Energie, D-14109 Berlin, Germany}

\author{Yi-De Chuang}
\affiliation{Lawrence Berkeley National Laboratory, Berkeley, CA 94720}

\author{Yu-Cheng Shao}
\affiliation{National Synchrotron Radiation Research Center,  101 Hsin-Ann Road, Hsinchu Science Park, Hsinchu, Taiwan 30076}

\author{Di-Jing Huang}
\author{Hsiao-Yu Huang}
\author{Amol Singh}
\affiliation{National Synchrotron Radiation Research Center, 101 Hsin-Ann Road, Hsinchu Science Park, Hsinchu, Taiwan 30076}

\author{Byron Freelon}
\affiliation{Physics Department and Texas Center for Superconductivity, University of Houston, Houston, TX 77204}

\date{\today}% It is always \today, today,

\begin{abstract}

Chromium tri-halides Cr$X_3$ ($X$ = Cl, Br, and I) have recently become a focal point of research due to their intriguing low-temperature, layer-dependent magnetic properties that can be manipulated by external stimuli. This makes them essential candidates for spintronics applications. Their magnetic orders are often related to the electronic structure parameters, such as spin-orbit coupling (SOC), Hund’s coupling ($J_H$), $p-d$ covalency, and inter-orbital Coulomb interactions. Accurately determining such parameters is paramount for understanding Cr$X_3$ physics. We have used high-resolution resonant inelastic x-ray scattering (RIXS) spectroscopy to study Cr$X_3$ across phase transition temperatures. Ligand field multiplet calculations were used to determine the electronic structure parameters by incorporating the crystal field interactions in a distorted octahedral orientation with $C_3$ symmetry. These methods provide the most detailed description of Cr$X_3$ magneto-optical and electronic energetic (terms) to date. The crystal field distortion parameters $D\sigma$ and $D\tau$ were experimentally determined, and the energies of $d$ orbitals have been reported. The spectroscopic measurements reveal an energy separation between spin-allowed quartet states and spin-forbidden doublet states, which increases upon going from CrCl$_3$ to CrI$_3$. The role of SOC in Cr $2p$ orbitals for the spin-flip excitations has been demonstrated. The determined 10$Dq$ values are in good agreement with the spectrochemical series, and Racah B follows the Nephelauxetic effect. Such precise measurements offer insights into the energy design of spintronic devices that utilize quantum state tuning within 2D magnetic materials.

\end{abstract}
\maketitle

\section{Introduction}

\begin{figure}[htbp]
     \centering
    \includegraphics[scale=1]{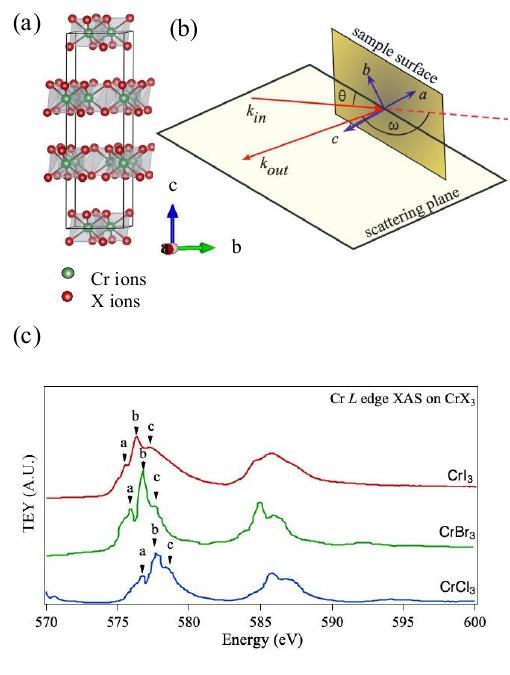}
    \caption{(a) Lattice structure of Cr$X_3$. The metal Cr and halide ions (X = Cl, Br, and I) are shown in green and red, respectively. (b) The schematic diagram illustrates the experimental setup for RIXS measurements. \textbf{$k_{in}$} and \textbf{$k_{out}$} are the incident and scattered photon beam wave vectors. (c) Cr $L$-edge XAS measurements on Cr$X_3$ ($X$ = Cl, Br, and I). The photon energies labeled \textbf{a, b} and \textbf{c} were used as the excitation energies in the RIXS measurements.}
    \label{fig: Experimental_setup}
\end{figure}

The chromium tri-halides Cr$X_3$ ($X$ = Cl, Br, and I) compounds have ferromagnetic (FM) order in the monolayer limit with Curie temperatures $T_c$ = 17 K, 34 K, and 61 K for bulk CrCl$_3$, CrBr$_3$, and CrI$_3$, respectively \cite{mondal2020magnetic,tsubokawa1960magnetic,mcguire2015coupling}. The interlayer interactions in bulk CrCl$_3$ are antiferromagnetic (AFM), while CrBr$_3$ and CrI$_3$ exhibit FM order. The electronegativity between metal Cr and the halides decreases as the halogen changes from Cl to Br to I \cite{shao2021spectroscopic,craco2021electronic,boeyens2008periodic}. This results in an increase in $p-d$ covalency from CrCl$_3$ to CrI$_3$. With the presence of spin-orbit coupling (SOC), the FM superexchange interaction across the $\sim 90^0$ Cr-$X$-Cr bonds becomes highly anisotropic in the edge-shared octahedral orientation in Cr$X_6$ (see Figure \ref{fig: Experimental_setup}(a)) \cite{goodenough1963magnetism}. This magnetocrystalline anisotropy stabilizes the magnetic order in Cr$X_3$, overcoming the Mermin-Wagner theorem \cite{mermin1966absence,ghosh2023magnetic}, and leads to rich phase diagrams in Cr$X_3$. This magnetic ordering appears to be dependent on the dimensionality, the atomic halogen constituents, layer interactions, and temperature. The electronic structure in the 2D magnetic materials reveals a strong interplay between the material's atomic structure, spin interaction, and the movement of electrons, which are crucial for applications in spintronics and data storage \cite{atanasov2012modern,craco2021electronic}. Understanding magnetism also appears to be linked to determining the electronic levels of chromium bands near the Fermi energy. These circumstances suggest that obtaining precisely determined energy scales is a prerequisite for constructing theoretical models that explain the magnetic ground states of Cr$X_3$. %Therefore, understanding the electronic structure of different halogen systems can provide a fundamental picture of the complex electric and magnetic properties of Cr$X_3$ and form a prerequisite for constructing theoretical models that explain the magnetic ground states of Cr$X_3$. Our ability to broadly exploit these promising materials in novel devices is limited by a lack of detailed knowledge of the underlying electronic structures, both theoretically and experimentally. 
The development of a high-accuracy Hamiltonian for Cr$X_3$, supported by experimental findings, is urgently needed to enhance the ability of the condensed matter physics community to utilize these intriguing materials. These findings could impact the field of spin-based electronics since magnetism in Cr$X_3$ compounds may be exploitable in the next generation of smart electronic devices, potentially surpassing the capabilities of graphene \cite{jiang2018controlling,novoselov20162d,pan2018large,akram2021moire,singamaneni2020light,mishra2024magnetic,ahmad2020pressure}.

Measurements, including infrared spectroscopy, optical absorption spectroscopy, Raman spectroscopy, and X-ray photoemission spectroscopy (XPS), have been identified as excellent techniques for probing electronic transitions and determining the electronic structure of the transition metal (TM) halides  \cite{pollini1970intrinsic,dillon1963magneto,limido1972specific,bermudez1979spectroscopic}. However, a significant challenge arises from the excitations between  $d$ orbitals being dipole forbidden \cite{suganotanabe,katsnelson2010theory}. Even though some $dd$ transitions can be observed with low intensity, electronic transitions with different spin multiplicities remain optically undetectable. This underscores the necessity for spectroscopic probes that detect all potential $dd$ transitions in order to determine a more accurate electronic structure in TM halides. 
With the development of synchrotrons, x-ray absorption spectroscopy (XAS), along with the resonant inelastic x-ray scattering (RIXS), has been used to probe the $dd$ and charge-transfer (CT) excitations in TM complexes \cite{Singh:yi5098,ament2011resonant,shao2021spectroscopic,occhialini2025spin,ghosh2023magnetic,BAKER2017182}. Cr $L_3$ edge RIXS is a resonant two-photon process ($2p$ $\xrightarrow{}$ $3d$ followed by $3d$ $\xrightarrow{}$ $2p$) that offers more flexibility in terms of selection rules compared to optical spectroscopy, primarily due to the core-hole intermediate state and its associated dynamics and interactions. The strong $2p$ core hole SOC permits the spin-selection rule to be alleviated, allowing for the probing of transitions that are spin-forbidden in optical spectroscopy. RIXS has been demonstrated to be a suitable tool for providing detailed electronic structure information in Cr$X_3$, with the state- and element-selective measurement capability providing detailed investigations in atomic, crystal ﬁeld, and charge-transfer information \cite{shao2021spectroscopic,ghosh2023magnetic,wang2017charge}.

This paper reports enhanced RIXS spectral features in Cr$X_3$ using high resolution spectroscopic measurements. A significant outcome of our study is the first experimental observation of a considerable energy separation between the spin-allowed quartet and spin-forbidden doublet states in Cr$X_3$. This energy gap shows an increment as the halogen changes from Cl to I. The Tanabe-Sugano-like energy level diagrams (ELDs) have been calculated and compared with the experimental RIXS spectra to determine the energy scales, such as crystal field splitting and Racah parameters. Atomic multiplet $2p-3d$ RIXS calculations have been performed to reconstruct the experimental RIXS spectra, and the electronic structure parameters across all halide systems in Cr$X_3$ have been summarized.

\section{Method}

\subsection{Experimental Method}

We performed Cr $L$-edge high resolution RIXS measurements on Cr$X_3$ at the 41A RIXS beamline of the Taiwan Photon Source (TPS) \cite{Singh:yi5098}. Figure \ref{fig: Experimental_setup}(b) illustrates the scattering geometry of RIXS measurements. The crystalline $c$-axis was aligned within the horizontal scattering plane ({$\pi$}-polarization) during the measurements. The $ab$ plane was unaligned, and the momentum dependence was not considered in this experiment. Cr $L$-edge XAS data were acquired prior to the RIXS measurements to determine the excitation photon energies. For the RIXS measurements, the tuned incident soft x-rays were directed at an angle ($\theta$) of 90$^{\text{o}}$ to the sample surface (normal incidence), and the spectrometer was set at 140$^{\text{o}}$ ($\omega$) back-scattering geometry with respect to the incoming photon beam. The instrumental energy resolution was {$\sim$30} meV. 

Complementary XAS and RIXS measurements were performed at the PEAXIS beamline, BESSY II, Germany, with a resolution of 120 meV. Cr$X_3$ single crystals were commercially obtained from HQ Graphene, and their crystallinity was verified using lab-based x-ray diffraction. Due to the samples' high hygroscopicity and sensitivity to oxygen, the Cr$X_3$ samples were stored and handled under an inert gas atmosphere (Ar) in a glovebox environment to minimize air exposure. The samples were subjected to scotch tape exfoliation before being transferred to the experimental chamber to ensure a clean surface.

\subsection{Computational Simulations}

The quantum many-body script language QUANTY \cite{haverkort2016quanty,crispy} was used to simulate the experimental Cr$X_3$ XAS and RIXS spectra. The Hamiltonian that describes the electronic structure of the system was built using Multiplet ligand field theory (MLFT) \cite{haverkort2016quanty,haverkort2012multiplet,de2008core}. The relevant electronic configurations in Cr$^{3+}$ for the RIXS process consist of a ground state $2p^63d^3$ with an intermediate excited state $2p^53d^4$ followed by a de-excitation to the ground state. They resulted in atomic multiplets, which are described by $3d-3d$ Coulomb and $2p-3d$ exchange interactions parameterized in Slater-Condon integrals $F^k_{dd}$, $F^k_{pd}$ (Coulomb), and $G^k_{pd}$ (Exchange) for Hatree-Fock calculations \cite{haverkort2016quanty,hunault2018direct}. The Racah B and Racah C parameters were used to account for the ion covalency, which can be related to the $F^2_{dd}$ and $F^4_{dd}$ by Racah B $= (9F^2_{dd} - 5F^4_{dd})/441$ and Racah C = 5$F^4_{dd}/63$ \cite{racah1942theory}. %3$d$ SOC was included to facilitate all the dipole and spin-forbidden transactions within RIXS simulations. 
The ligand-to-metal charge transfer (LMCT) parameters were incorporated to account for hopping between the halide ions and Cr$^{3+}$ metal ions.

Local symmetry is a crucial parameter that can significantly influence the electronic, magnetic, and optical properties of materials \cite {hunault2016effect}. While the perfect octahedral ($O_h$) symmetry has been commonly assumed for the metal Cr$^{3+}$ in Cr$X_3$ studies \cite{shao2021spectroscopic,seyler2018ligand,dillon1963magneto,craco2021electronic,he2025dispersive}, only a few have accounted for the actual distorted $O_h$ orientation. The trigonal distortion in Cr$X_3$ results in lowering the symmetry from $O_h$ to $D_{3d}$ and further to $C_3$ \cite{atanasov2012modern,ghosh2023magnetic,DILLON19661531,pollini1970intrinsic,occhialini2025spin,bermudez1979spectroscopic}. Notably, $C_3$ symmetry is a subgroup of the $D_{3d}$ symmetry \cite{juhin2008x}. Since the high resolution RIXS spectra enabled us to resolve multiple peak splittings, atomic multiplet RIXS spectra calculations using lower symmetry $C_3$ reproduced the experimental RIXS spectra more reliably.

\begin{figure*}[!htbp]
    \centering
    \includegraphics[scale=1]{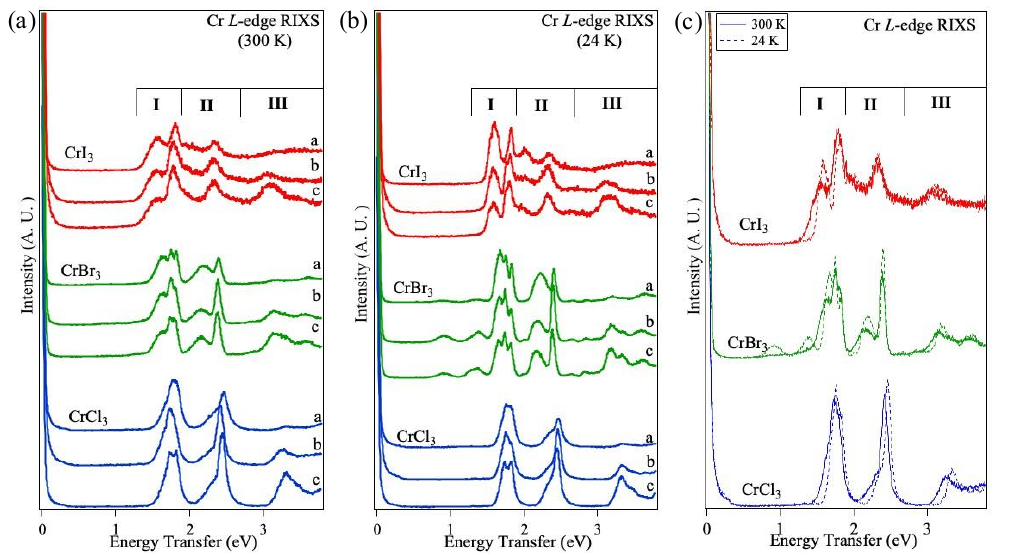}
    %{\includegraphics[width=1\textwidth]{RIXS_CrX3_tth140_th90.pdf}}
    \caption{ (a) Cr $L_3$-edge RIXS data measured in Cr$X_3$ at 300 K and (b) at 24 K. The RIXS data have been collected at three different excitation energies: \textbf{a, b,} and \textbf{c} (see Figure \ref{fig: Experimental_setup}(c)). The three regions, I, II, and III, show different spectral features in RIXS spectra. (c) Temperature comparison of the RIXS data at Cr $L_3$ edge. Solid (dashed) lines indicate the 300 K (24 K) RIXS data.}
    \label{fig: RIXS_CrX3_tth140_th90_pi}
\end{figure*}

\section{Results}

\subsection{XAS and RIXS}

Figure \ref{fig: Experimental_setup}(c) shows the Cr $L$-edge XAS spectra acquired from 570 to 600 eV at room temperature (RT) in total electron yield (TEY) mode. During the XAS process, the $2p^63d^3$ ground state electrons are excited to the $2p^53d^4$ state. Therefore, spectral features are dominated by dipole transitions from the core $2p$ level to the empty $3d$ states. Consequently, the two primary peaks observed in the XAS data can be attributed to the $L_3$ ($\sim$575 - 580 eV) and $L_2$ ($\sim$583 - 590 eV) lines \cite{serri2020enhancement,shao2021spectroscopic}. 
The $L_3$-edge (\textbf{b} in Figure \ref{fig: Experimental_setup}(c)) in Cr$X_3$ was determined to be 577.7 eV, 576.8 eV, and 576.4 eV for $X$ = Cr, Br, and I, respectively; indicating a shift towards lower energy as the halogen changes from Cl to I.

Next, the incident photon beam was tuned to energies \textbf{a, b,} and \textbf{c}, and the RIXS measurements were performed. Figure \ref{fig: RIXS_CrX3_tth140_th90_pi}(a) shows the experimental RIXS spectra measured at RT. The 0 eV feature can be attributed to the elastic feature, where the incident photon energy $E_{in}$ equals the scattered photon energy $E_{out}$. The nonzero spectral features are divided into three regions: I (1.3 - 1.9 eV), II (1.9 - 2.7 eV), and III (2.7 - 4 eV). %All RIXS spectra were normalized to the main peak intensity in Region I to facilitate comparison. 
All three regions, I, II, and III, contain spectral features that can be attributed to the inter-orbital $dd$ excitations. In this report, we only focus on $d$-site electronic structure and related excitations. Spectral features related to the CT process above 4 eV will be discussed elsewhere. 

In Figure \ref{fig: RIXS_CrX3_tth140_th90_pi}(a), CrCl$_3$ shows a broader single peak in region I, while CrBr$_3$ shows a shoulder feature, and CrI$_3$ reveals a distinct splitting between the two peaks. This splitting becomes more pronounced from Cl to I. In region II, CrCl$_3$ has a main peak with a shoulder characteristic, and CrBr$_3$ shows a clear separation between two peaks, while the shoulder characteristic in CrI$_3$ is dominated by the main peak in region I. Region III consists of more broadened multiple subpeaks in all three materials. The energy transfer values in the aforementioned $dd$ regions are independent of the incident energies and are known as Raman-like losses. However, the spectral intensities change with the excitation energies (\textbf{a} to \textbf{c}). These resonant and Raman behaviors further justify the attribution of these peaks to the well-known $dd$ excitations.

Since Cr$X_3$ shows structural and magnetic phase transitions at different temperatures, we repeated the RIXS measurements at the low temperature (LT) of 24 K (see Figure \ref{fig: RIXS_CrX3_tth140_th90_pi}(b))-(c). At 24 K, CrBr$_3$ and CrI$_3$ are magnetically ordered, whereas CrCl$_3$ remains paramagnetic. The same energy regions are demarcated in the LT RIXS spectra, and the spectral features are much sharper. In particular, the intensity of the shoulder features increases at 24 K. A peak shift is also observed: in CrBr$_3$, the energy gap between the shoulder feature and main peak in region II becomes smaller at LT. Moreover, several extraneous peaks emerge in LT RIXS spectra of CrBr$_3$ between 0 and 1.5 eV. However, such features are absent in the LT RIXS data from BESSY II and ALS \cite{shao2021spectroscopic}. Therefore, a more extensive analysis is required to validate these LT features in the 0 - 1.5 eV range, and their physical origin remains unresolved.  %CrI$_3$ displays well-separated peaks in the regions I and II at LT.%Appendix \ref{RIXS spectral features in 0 - 1.5 eV} discusses a brief comparison of these low energy features. %, and we suggest that momentum-resolved RIXS might enable a better determination of these features.
%Next, we move to the RIXS spectra simulations to understand the nature and the electronic excitations in the experimental RIXS data.

\subsection{Energy Level Diagrams and RIXS Simulations}

\subsubsection{Energy Level Diagrams}

In this study, electronic ELDs are utilized to analyze the spectral features observed in the experimental RIXS data. The ELDs are calculated by diagonalizing the standard ligand field multiplet Hamiltonian. The results provide the number of excited states in the system, including the spin-allowed and spin-forbidden states, which can be directly compared with the RIXS spectral features to determine the corresponding energy scales.

\begin{figure}[ht]
    %\centering
    %{\includegraphics[width=0.5\textwidth]{d_levels.pdf}}
    {\includegraphics[scale=1]{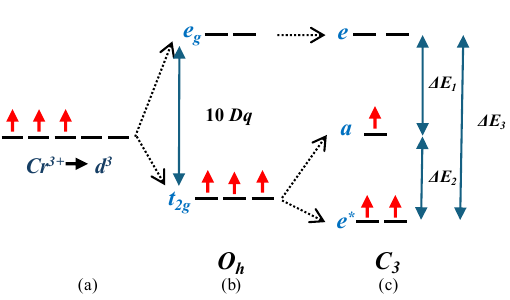}}
    \caption{(a) valence $d$ orbital electron distribution in Cr$^{3+}$ metal ion. (b) Lifting of degeneracy of the $d^{3+}$ spectroscopic term (free $F^{4+}$ ion) due to $O_h$ symmetry. This configuration’s 5 $d$ orbitals are divided into two energy levels $t_{2g}$ and $e_g$. (c) Lifting of the degeneracy of the $d^{3+}$ electrons due to $C_3$ symmetry. The five $d$ orbitals are divided into one $a$ state and two $e$ states within the $C_3$ symmetry.}
    \label{fig: d_level_splitting}
\end{figure}

The multielectronic atomic states of the Cr$^{3+}$($d^3$) configuration are labeled in the diagrams. In a perfect octahedral environment with $O_h$ symmetry, the Cr five $d$ orbitals are split into two energy levels $t_{2g}$ and $e_g$ (see Figure \ref{fig: d_level_splitting}(b)). This crystal field splitting results in the division of the $3d$ multielectronic atomic quartet states into manifolds: $^4F$ $\xrightarrow{}$ $^4A_{2g}$, $^4T_{2g}$, and $^4T_{1g}(1)$, and $^4P$ $\xrightarrow{}$ $^4T_{1g}(2)$ \cite{vercamer2016calculation,shao2021spectroscopic}. When the trigonal distortion is considered and the symmetry is lowered from $O_h$ to $C_{3}$ ($D_{3d}$), the $t_{2g}$ orbitals are further divided into two states $a$ ($a_{1g}$) and $e^*$ ($e_g^*$), as shown in Figure \ref{fig: d_level_splitting}(c). Within the $C_3$ symmetry configuration, the atomic state $^4T_{2g}$ and $^4T_{1g}$ branches into $^4A$ and $^4E$ states  \cite{DILLON19661531,occhialini2025spin,hunault2018direct}. The energies of the three states $e$, $a$, and $e^*$, can be written using the crystal field parameter $Dq$, and the distortion parameters $D\tau$, and $D\sigma$ as follows : \cite{juhin2008x,haverkort2016quanty}.

\begin{subequations}
\begin{eqnarray}
E_e = 6Dq + (7/3) D\tau \label{appa} \\
E_{a} = -4Dq -2 D\sigma - 6 D\tau \label{appb} \\
E_{e*} = -4Dq + D\sigma + (2/3) D\tau \label{appc}
\end{eqnarray}
\label{eq: d_level_energies}
\end{subequations}

\begin{figure*}[!htbp]
    \centering
    %{\includegraphics[width=0.5\textwidth]{ELD_CrCl3.pdf}}
    {\includegraphics[scale=1]{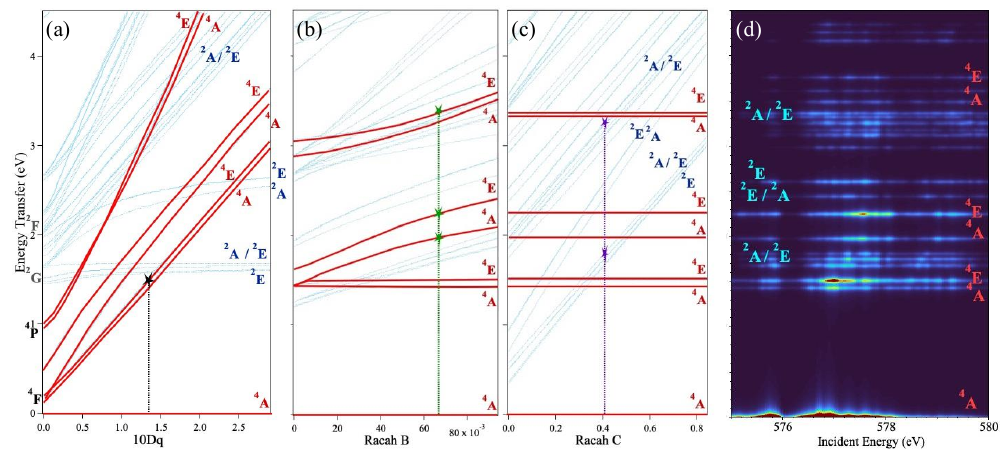}}
    \caption{ (a) Energy level diagrams as a function of crystal field $Dq$ (b) Racah B and (c) Racah C in CrI$_3$. The quartet and doublet states in Cr$^{3+}$ metal are shown by solid red and dashed blue lines, respectively. The black, green, and purple stars indicate the extracted $Dq$, Racah B, and Racah C values by comparing the ELDs with experimental RIXS spectra. (d) Calculated RIXS map recorded at 300 K. A broadening of 30 meV was considered, similar to the experimental resolution.}
    \label{fig: ELDs in RT CrI3}
\end{figure*} 

The initial parameters for ELD calculations were referred from the literature corresponding to the $O_h$ symmetry \cite{shao2021spectroscopic}. The Coulomb interaction $U$ of 3.5 eV was used, and the ELDs do not depend on the Coulomb interaction $U$, and its effect on the RIXS intensity is very small \cite{feldkemper1998generalized}. The ratio between $U_{dd}$ and $U_{pd}$ was kept constant at 1.5. Figure \ref{fig: ELDs in RT CrI3} shows the energy level diagrams calculated as a function of the crystal field $Dq$, Racah B, and Racah C in CrI$_3$. Dashed blue and solid red lines indicate the spin-doublet and spin-quartet states, respectively. Here we only report the ELDs for CrI$_3$, as it has the maximum splitting between energy states. 

The crystal field splitting 10$Dq$ can be determined by considering the electronic transition $^4A_{2g}$ $\xrightarrow{}$ $^4T_{2g}$ states within the $O_h$ environment \cite{dalal2017textbook}. This is equivalent to the first non-zero energy transfer peak maximum in experimental RIXS spectra, where the $^4T_{2g}$ state has been reported at 1.73 eV, 1.6 eV, and 1.45 - 1.58 eV for CrCl$_3$, CrBr$_3$, and CrI$_3$, respectively \cite{schlafer1969basic,shao2021spectroscopic,ghosh2023magnetic,occhialini2025spin}. These values align with the shoulder feature in region I in our RIXS data. On the other hand, the sharp peaks in the optical spectroscopy can be compared directly with the $dd$ excitations in the RIXS spectra. Optical absorption and magnetic circular dichroism (MCD) measurements report the spin allowed $^4T_{2g}$ states and bright excitons in CrCl$_3$ at 1.7 - 1.75 eV, CrBr$_3$ at 1.6 - 1.65 eV, and CrI$_3$ at 1.45 - 1.6 eV, respectively \cite{bermudez1979spectroscopic,pollini1989optical,pollini1970intrinsic,acharya2021excitons,dillon1963magneto,he2025dispersive,limido1972specific}. These peak positions also show good alignment with the shoulder features in region I. Therefore, considering the shoulder peak in region I and comparing that with the ELD in Figure \ref{fig: ELDs in RT CrI3}(a), the crystal field splitting 10$Dq$ in CrCl$_3$, CrBr$_3$, and CrI$_3$ was determined to be 1.58 eV, 1.54 eV, and 1.35 eV, respectively. The corresponding electronic transition is $^4A$ $\xrightarrow{}$ $^4E$ ($t_{2g}^3$ $\xrightarrow{}$ $t_{2g}^2e_g^1$) for the $C_3$ symmetry, since the transition of orbital singlets $^4A$ $\xrightarrow{}$ $^4A$ is dipole-forbidden. The splitting of $^4E$ and $^4A$ ($^4T_{2g}$) states can be tuned by the distortion parameter $D\sigma$, which was determined to be -20 meV, -15 meV, and -10 meV as the halogen changes from Cl to I. A summary of the crystal field parameters and the energies of each state is listed in Table \ref{tab: crystal field parameters}.

The Racah B was varied next, and the result is presented in Figure \ref{fig: ELDs in RT CrI3}(b). The two peaks in region II were assigned to $^4A$ and $^4E$ states, and they showed an increment in the splitting as the halogen changed from Cl to I. Their expected order of states and the energy gap were dependent on the crystal field distortion parameter $D\tau$, which was determined to be 27 meV, 33 meV, and 35 meV for CrCl$_3$, CrBr$_3$, and CrI$_3$, respectively. The electronic transition $^4A$ $\xrightarrow{}$ $^4E$ ($^4T_{1g}$) which corresponds to the double electron excitation $t_{2g}^3$ $\xrightarrow{}$ $t_{2g}^1e_g^2$ was considered in determining the Racah B values of 0.072 eV, 0.071 eV, and 0.067 eV in CrCl$_3$, CrBr$_3$, and CrI$_3$, respectively. 

After tuning the 10$Dq$ and Racah B values, the ELD was calculated as a function of Racah C, as shown in Figure \ref{fig: ELDs in RT CrI3}(c). At this stage, the quartet lines appear to be horizontal. However, the ELDs include several doublet states influenced by Racah B and Racah C. These doublet states represent spin-flip electronic transitions. The SOC in the Cr $2p$ level can accompany the spin-flip electronic transition during the RIXS process, producing spin-forbidden doublet states with $S$ = 1/2, as discussed in the latter. Racah C of 0.41 - 0.42 eV in all Cr$X_3$ was determined after considering the transitions $^4A$ $\xrightarrow{}$ $^2A$ and $^4A$ $\xrightarrow{}$ $^2E$ ($^2T_{1g}$). A summary of the energy scales calculated in Cr$X_3$ at RT is listed in Table \ref{tab: electronic structure parameters}.

\begin{table*}
\caption{\label{tab: crystal field parameters}Summary of the crystal field splitting parameters and the energy levels in Cr $d$ orbitals in Cr$X_3$.}
 \begin{ruledtabular}
        \begin{tabular}{ccccccccccc} 
            & T & 10$Dq$ & $D\tau$ & $D\sigma$ & $E_e$ &  $E_{a}$& $E_{e*}$&$\Delta$$E_1$& $\Delta$$E_2$&$\Delta$$E_3$\\ \hline 
            CrCl$_3$ & 300 K & 1.580 & 0.027 & -0.020 & 1.011 & -0.754 & -0.634 & 1.765&-0.120&1.645\\
            CrBr$_3$ & 300 K & 1.540 & 0.033 & -0.015 & 1.001 & -0.784 & -0.609 &1.785&-0.175&1.610\\
            CrI$_3$ & 300 K & 1.350 & 0.035 & -0.010 & 0.892 & -0.730 & -0.527 &1.622&-0.203&1.418\\
            & & & & & \\
            CrCl$_3$ & 24 K & 1.620 & 0.020 & -0.020 & 1.019 & -0.728 & -0.655 & 1.747&-0.073&1.673\\
            CrBr$_3$ & 24 K & 1.540 & 0.028 & -0.015 & 0.989 & -0.754 & -0.612 &1.743&-0.142&1.602\\
            CrI$_3$ & 24 K & 1.370 & 0.035 & -0.010 & 0.904 & -0.738 & -0.535 &1.642&-0.203&1.438\\
        \end{tabular}
    \end{ruledtabular}
\end{table*}

\subsubsection{XAS/RIXS spectrum calculation}

\begin{figure*}
    \centering
    {\includegraphics[scale=1]{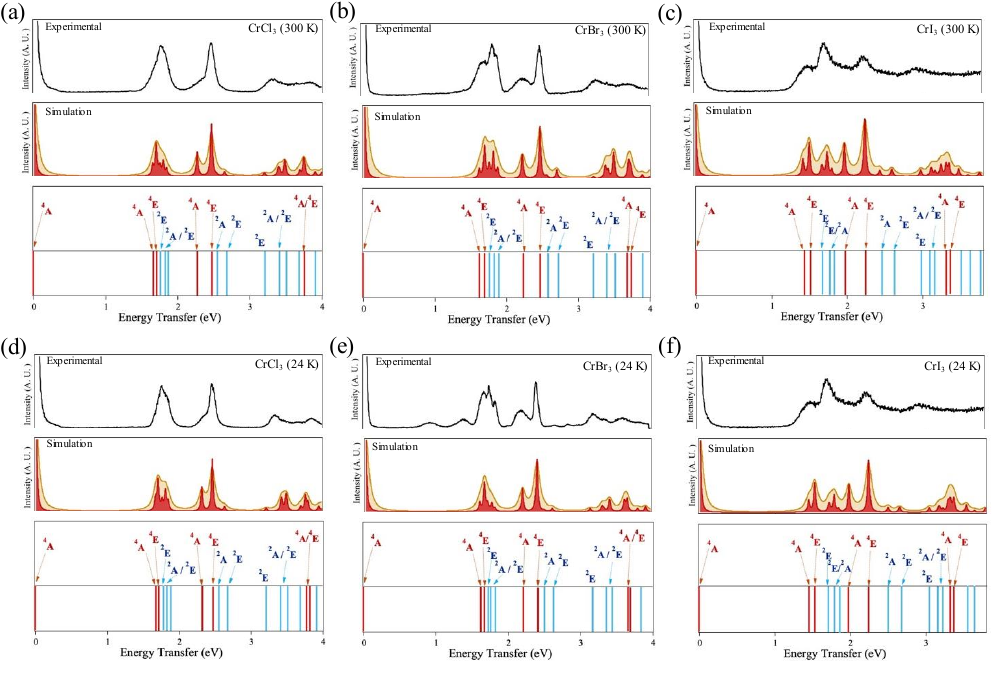}}
    \caption{Comparison of experimental and simulated RIXS spectra at Cr $L_3$ edge (a)-(c) at 300 K and (d) - (f) at 24 K in CrCl$_3$, CrBr$_3$, and CrI$_3$, respectively. In each figure, the top panel and the middle panels show the experimental and simulated RIXS spectra at the Cr $L_3$ edge. Bottom panel indicates the spectroscopic term labels given in $C_3$ symmetry, analogous to the middle panel.}
    \label{fig: RIXS_fitting_results_CrX3}
\end{figure*}

The XAS experimental data were used to tune the RIXS intermediate parameters, as the RIXS process consists of an XAS process followed by a resonant X-ray emission. The Slater integral parameters $F^2_{pd}$, $G^1_{pd}$, and $G^3_{pd}$, which account for the electron exchange interactions $2p3d$, were tuned to match the experimental XAS data. Furthermore, the $2p_{3/2}$ and $2p_{1/2}$ spectral parts are clearly separated by the core-hole SOC and the core-hole lifetime broadening, which gives the sharp features \cite{van19922p}. The energy separation of 8.5 eV between the $L_3$ and $L_2$ peaks (Figure \ref{fig: Experimental_setup}(c)) suggested a Cr $2p$ SOC value of $\sim$5.7 eV. 

Subsequently, we calculated the RIXS spectra using all refined parameters and compared the results with the experimental RIXS data, as illustrated in Figure \ref{fig: RIXS_fitting_results_CrX3}.  In each figure, the top panel shows the experimental RIXS data measured at the Cr $L_3$ edge. The middle panel shows the calculated RIXS spectra analogues of the top panel. The RIXS spectra calculated with a broadening of 30 meV (red shaded line) exhibit fine spectral features and energy positions that are responsible for the main intensities in the experimental data. We only reported the results at the $L_3$-edge (excitation energy \textbf{b}) in Figure \ref{fig: RIXS_fitting_results_CrX3}. Similar results were obtained for all the excitation energies (see the RIXS map in Figure \ref{fig: ELDs in RT CrI3}(c)). The peak positions extracted from the simulated RIXS spectra are shown in the bottom panel with red lines indicating spin-allowed quartet states and blue lines indicating spin-flip doublet states. Figure \ref{fig: RIXS_fitting_results_CrX3} clearly demonstrates the RIXS ability to probe states with diﬀerent spin multiplicities selectively.

A clear energy gap between the spin-allowed quartets and spin-forbidden doublets in regions I and II was observed with high resolution in RIXS data. This energy difference becomes more prominent as the halogen changes from Cl to Br to I, which explains the appearance of a shoulder feature as the halogen changes from Cl to I in the experimental RIXS data (see Figure \ref{fig: RIXS_fitting_results_CrX3}). Two peaks observed in region II primarily depend on the crystal field distortion parameter $D\tau$ as they can be attributed to $^4A$ and $^4E$ states and result in improved RIXS fitting using the distorted octahedral orientation with $C_3$ symmetry in Cr$X_3$. There is a slight mismatch between the experimental RIXS data and the simulated RIXS spectra in region III due to the cluster of doublets (see Figure \ref{fig: ELDs in RT CrI3}) and CT excitations, which we haven't discussed in this study. As the temperature is lowered to 24 K, the shift in the RIXS spectral features in region I (see Figure \ref{fig: RIXS_CrX3_tth140_th90_pi}(c)) results in higher $10Dq$ values and thus a greater crystal field splitting at LT. This increase in splitting will enhance the single-ion magnetic anisotropy, which helps stabilize the magnetic order at LT, requiring more energy to probe the $t_{2g}$ $\xrightarrow{}$ $e_g$ excitation  \cite{zheng2011theoretical,kvashnin2020relativistic}. In region II, the shift in CrI$_3$ is negligible. The peak shift in CrCl$_3$ and CrBr$_3$ can be explained by $D\tau$ and Racah B. Since Cr$X_3$ exhibits both MPT and SPT at different temperatures, the SPT can possibly cause a change in octahedral distortion, resulting in different crystal field distortion parameters and, consequently, different electronic structure parameters.

\begin{figure}
    \centering
    %{\includegraphics[width=0.5\textwidth]{spin_excitations.pdf}}
    {\includegraphics[scale=1]{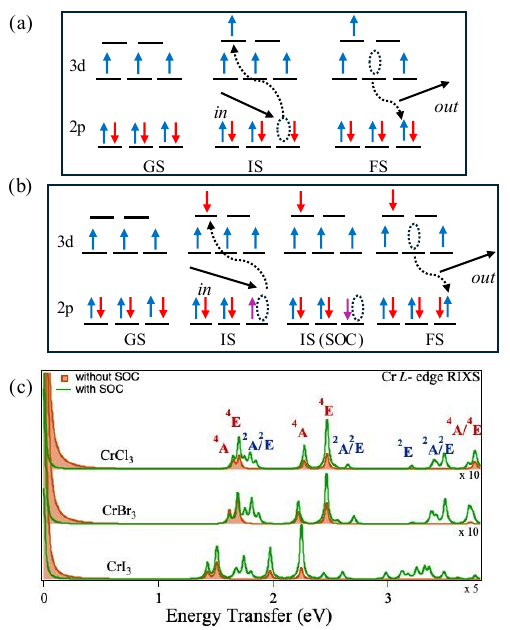}}
    \caption{(a) Electronic excitation corresponds to the spin-allowed quartet state. The ground state (GS), intermediate state (IS), and final state (FS) electron configurations are indicated. The spin moment is conserved throughout the process. (b) Electronic excitation for the spin-flip doublet state. The incident and scattered x-rays are labeled as $in$ and $out$, respectively. (c) Comparison of RIXS spectra with the contribution of SOC (green) and without SOC (brown) in the Cr metal ion.}
    \label{fig: spin_excitations_CrX3}
\end{figure}

A schematic representation of Cr $2p3d$ RIXS in Figure \ref{fig: spin_excitations_CrX3} (a),(b) shows the electronic excitations from the initial ground state through the intermediate core-excited to the ﬁnal valence-excited states. Figure \ref{fig: spin_excitations_CrX3} (a) captures the electronic excitation corresponding to the spin-allowed quartet states, where the total spin moment of the ground state and the final state is $S=3/2$ ($2S+1 = 4$), indicating a conserved spin moment during the RIXS process. On the other hand, Figure \ref{fig: spin_excitations_CrX3} (b) shows the electronic excitation that corresponds to the spin-forbidden (spin-flip) doublet states, where the total spin moment changes from $S=3/2$ to $S=1/2$ ($2S+1 = 2$) from the ground state to the final state, indicating a spin-flip during the process. The presence of Cr $2p$ SOC (given by the Cr $2p$ spin-orbit parameter $\zeta(2p)$ (5.668 eV)) which is much stronger than the $3d$ SOC (given by the Cr $3d$ spin-orbit parameter $\zeta(3d)$ (0.018 eV)) contributes to this spin-flip process and enables the probing of spin-forbidden states \cite{van19922p,dillon1963magneto}. Figure \ref{fig: spin_excitations_CrX3} (c) compares the RIXS spectra calculated with the SOC (green line) and without SOC (brown curve). Spectra without SOC do not produce the spin-forbidden doublet states, disagreeing with the experimental data. Therefore, our calculation suggests that SOC must be considered in RIXS calculations in order to probe the spin-forbidden doublet states.

\section{Discussion}

Based on high resolution RIXS spectra for Cr$X_3$, a detailed electronic structure parameter calculation was presented, including the determination of the crystal field splitting and Racah parameters in Cr$X_3$. Now, we discuss how the obtained values relate to previously determined electronic structure parameters in Cr$X_3$. The extracted energies of the spin-quartet and spin-doublet states are consistent with the $d^3$ ELDs in Cr$^{+3}$ systems \cite{shao2021spectroscopic,hunault2018direct}. Density functional theory calculations predict that the magnetic moment is hosted predominantly in the Cr$^{3+}$ ions with $S$ = 3/2 spin configuration and a reported spin moment of 3$\mu_B$/Cr \cite{molina2020magneto,mcguire2015coupling,dillon1965magnetization}. From our calculations, the spin moment of the Cr was determined to be $\sim$3.75 $\mu_B$/Cr, indicating a larger value compared to the nominal spin moment in Cr due to the ligand contribution \cite{shao2021spectroscopic,liu2023light}.

\begin{table*}
\caption{\label{tab: electronic structure parameters}Summary of the electronic structure parameters calculated at 300 K for Cr$X_3$.}
 \begin{ruledtabular}
        \begin{tabular}{ccccccc} 
            & CrCl$_3$ & CrCl$_3$ & CrBr$_3$ & CrBr$_3$ & CrI$_3$ & CrI$_3$\\
            & $(3d^3)$ & $(2p^53d^4)$ & $(3d^3)$ & $(2p^53d^4)$ & $(3d^3)$ & $(2p^53d^4)$ \\ \hline
            Racah B & 0.072 & 0.077 & 0.071 & 0.076 & 0.067 & 0.072\\
            Racah C & 0.410 & 0.441 & 0.420 & 0.452 & 0.410 & 0.441\\
            $F^2_{dd}$ & 6.398(59\%) & 6.884(59\%) & 6.419(60\%) & 6.907(60\%) & 6.153(57\%) & 6.620(57\%)\\
            $F^4_{dd}$ & 5.166(76\%) & 5.561(76\%) & 5.292(78\%) & 5.696(78\%) & 5.166(76\%) & 5.561(76\%)\\
            $F^2_{pd}$ & - & 5.873(90\%) & - & 5.546(85\%) & - & 5.220(90\%)\\
            $G^1_{pd}$ & - & 3.254(68\%)& - & 3.110(65\%) & - & 2.871(60\%)\\
            $G^3_{pd}$ & - & 1.633(60\%) & - & 1.361(50\%) & - & 1.088(40\%)\\
            SOC$_{3d}$ & 0.018(50\%) & 0.018(50\%) & 0.018(50\%) & 0.018(50\%) & 0.018(50\%) & 0.018(50\%)\\
            SOC$_{2p}$ & - & 5.668(100\%) & - & 5.668(100\%) & - & 5.668(100\%) \\
            $J_H$ & 0.826 & 0.889 & 0.837 & 0.900 & 0.809 & 0.870 \\
            $\Delta$ & 3.800 & 3.800 & 3.300 & 3.300 & 3.000 & 3.000 \\
            $10DqL$ & 0.103 & 0.103 & 0.238 & 0.238 & 0.032 & 0.032 \\
            $V_{e}$ & 1.992 & 1.992 & 1.940 & 1.940 & 1.836 & 1.836 \\
            $V_{a1}$ & 1.320 & 1.320 & 1.380 & 1.380 & 1.400 & 1.400 \\
            $V_{e*}$ & 1.320 & 1.320 & 1.380 & 1.380 & 1.400 & 1.400 \\
            % & & & \\
        \end{tabular}
    \end{ruledtabular}
    \text{*All values are in eV}
\end{table*}

The reported 10$Dq$ values in CrCl$_3$ (1.50 - 1.75 eV), CrBr$_3$ (1.45 - 1.65 eV), and CrI$_3$ (1.30 - 1.45 eV) exhibit significant overlap and variability, indicating inconsistency and a varience in the determined values \cite{shao2021spectroscopic,occhialini2025spin,pollini1970intrinsic,bermudez1979spectroscopic,ghosh2023magnetic,dillon1963magneto}. This makes it hard to draw firm conclusions about the trend in the ligand field strength. The improved energy resolution from our RIXS measurements enabled further refinement of the $10Dq$ parameter using $C_3$ symmetry with distorted octahedral orientation across all halides. As the halogen changes from Cl to Br to I, a clearly decreasing trend is observed in the refined $10Dq$ values from our study (1.58 eV, 1.54 eV, and 1.35 eV for CrCl$_3$, CrBr$_3$, and CrI$_3$, respectively). This is in good agreement with the spectrochemical series explaining the strength of various ligand-induced crystal fields: I$^{-1} <$ Br$^{-1} <$ Cl$^{-1}$ \cite{tchougreeff2009nephelauxetic}. This work highlights the need for high-resolution RIXS measurements when it comes to determining precise electronic structure parameters.

\begin{figure}
    \centering
    {\includegraphics[scale=1]{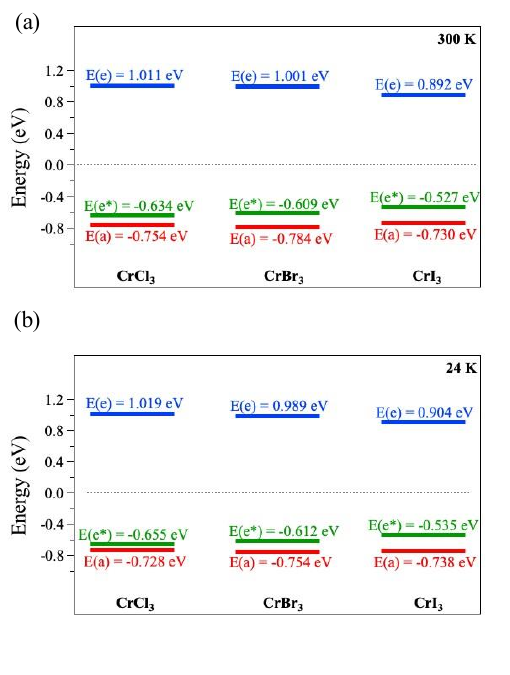}}
    \caption{ The calculated energy levels of the Cr $d$ orbitals in CrCl$_3$, CrBr$_3$, and CrI$_3$ at (a) 300 K and (b) 24 K.}
    \label{fig: Energy levels in CrX3}
\end{figure}

The determined crystal field distortion parameters (see Table \ref{tab: crystal field parameters}) show an opposite sign relation: negative for $D\sigma$ and positive for $D\tau$. As the halogen changes from Cl to I, the $D\sigma$ and $D\tau$ show incremental changes, suggesting a higher (lower) distortion in CrI$_3$ (CrCl$_3$). A similar study reports the distortion parameters for CrI$_3$: $D\sigma$ = -0.3 eV and $D\tau$ = 0 eV, and CrF$_2$: $Ds$ = -0.2 eV and $Dt$ = 0.2 eV, showing a good agreement with our results \cite{ghosh2023magnetic,jimenez2015x}. The determined crystal field parameters were applied to the equations (\ref{appa}) - (\ref{appc}) to calculate the energies of the states $e$, $a$, and $e*$ as summarized in Table \ref{tab: crystal field parameters} and graphically represented in Figure \ref{fig: Energy levels in CrX3}. The results suggest $a$ state has the lowest energy. A similar 2D van der Waals family V$X_3$ reports a trigonal contraction when the $a$ state is lowest, and our results suggest a similar behaviour in Cr$X_3$ \cite{sant2023anisotropic,camerano2024symmetry}. The energy gap between each state is given by $\Delta$$E_1$, $\Delta$$E_2$, and $\Delta$$E_3$ (see Figure \ref{fig: d_level_splitting}) as summarized in Table \ref{tab: crystal field parameters}. The energy gap between the highest and lowest states ($\Delta$$E_3$) decreases as the halogen changes from Cl to I, consistent with the $10Dq$ trend.

A summary of the electronic structure parameters used in the RIXS simulations at 300 K is listed in Table \ref{tab: electronic structure parameters}, including the LMCT parameters. Note that the unequal scaling in Hatree-Fock Slater integrals $F^2_{dd}$ and $F^4_{dd}$, and the exchange parameters $G^1_{pd}$ and $G^3_{pd}$ are essential to reproduce the RIXS spectra. The Racah B parameter, which accounts for inter-electronic repulsion, shows a decreasing trend as the halogen changes from Cl to Br to I, agreeing with the nephelauxetic effect. These values reflect the increasing covalency that is observed from Cl $<$ Br $<$ I with higher scaling of the Hatree-Fock parameters corresponding to the ionic interactions and lower scaling corresponding to the covalent interactions \cite{shao2021spectroscopic,tchougreeff2009nephelauxetic,atanasov2012modern,haverkort2012multiplet}. The determined Racah B values show a good agreement with the reported values, which are in range of 0.067 - 0.095, 0.062 - 0.09, and 0.047 - 0.087 for CrCl$_3$, CrBr$_3$, and CrI$_3$, respectively \cite{shao2021spectroscopic,atanasov2012modern}. The Racah C values do not change as the halogen varies. The Racah A (A $= F^0_{dd} - F^4_{dd}/9$ )was ignored because it is considered the same for any metal ion. The determined C/B ratios at RT of 5.69, 5.92, and 6.12 were lower than the literature-reported values and closer to the expected value of (3.75 - 4). The Racah B parameter primarily determines the ratio C/B, as the Racah C is constant. It increases with the halogen change from Cl to I, reflecting the shift from ionic bonding in CrCl$_3$ to covalent bonding in CrI$_3$. The Hund’s exchange coupling $J_H = (F^2_{dd} + F^4_{dd})/14$ for CrCl$_3$, CrBr$_3$, and CrI$_3$ was calculated to be 0.826, 0.837, and 0.809 eV, respectively \cite{shao2021spectroscopic,craco2021electronic}. These values indicate that all Cr$X_3$ materials are in the high spin (HS) state, given the relation 2$J_H$ $<$ 10$Dq$ \cite{tomiyasu2017coulomb}.

The SOC in Cr$X_3$ plays a significant role in the spin-flip (spin-forbidden) process, as discussed earlier. Recent studies suggest that the spin-flip excitations are responsible for the anisotropic SOC, which, in turn, contributes to the Dzyaloshinskii-Moriya (DM) interaction \cite{miyawaki2017dzyaloshinskii,ghosh2023magnetic}. However, further investigations are needed to better understand the relationship between the strength of SOC and its influence on the DM interaction. This study lays an important foundation for future investigations into Cr$X_3$ as a material for spintronic and optoelectronic devices, where precise control over electronic and optical properties is crucial for achieving advanced functionalities. 

\section{Conclusions}

We conducted Cr $2p-3d$ XAS and Cr $L$-edge high resolution RIXS measurements on Cr$X_3$. Atomic multiplet RIXS simulations, interpreted using the MLFT approach, showed good agreement with the experimental RIXS data, as illustrated in Figure \ref{fig: RIXS_fitting_results_CrX3}. This study marks the first experimental observation of clear energy separations between spin-allowed and spin-forbidden electronic transitions in Cr$X_3$ using the improved high-resolution RIXS measurements. By applying ELDs in atomic multiplet calculations, we have refined the electronic energy scales, taking into account the distorted octahedral orientation with $C_{3}$ symmetry in Cr$X_3$, as detailed in Table \ref{tab: crystal field parameters} and \ref{tab: electronic structure parameters}. These findings indicate that the ligand substitution significantly affects the $d$-orbital electronic structure. The superexchange coupling effect on the inter-atomic orbital splitting was discussed, as well as the DM interactions and the role of SOC in the spin-forbidden excitations. We propose that investigating RIXS on few-layer Cr$X_3$ samples could offer deep insights into the electronic structure properties of these materials. Further investigations are needed to understand the low-energy RIXS spectral features in the 0 - 1.5 eV range. 

\begin{acknowledgments}

This research was conducted at 41A RIXS beamline, Taiwan Photon Source (TPS), and PEAXIS RIXS beamline, BESSY II, Germany operated by the Helmholtz-Zentrum Berlin für Materialien und Energie. The preliminary RIXS measurements were taken at the qRIXS 8.0.1 beamline, Advanced Light Source (ALS), which is a DOE Office of Science User Facility under contract no. DE-AC02-05CH11231. We thank all the beamline scientists for their valuable support. Y.C.S. acknowledges the financial support from the National Science and Technology Council (NSTC) in Taiwan under grant numbers 113-2112-M-213-025-MY3. The Welch Foundation (grant number: E-0001) and the Texas Center for Superconductivity (TCSUH) supported work at the University of Houston. Part of this work was supported by the U.S. DOE, BES, under Award No. DE-SC0024332. The authors acknowledge support from the U.S. Air Force Office of Scientific Research and Clarkson Aerospace Corp. under Award FA9550-21-1-0460. Special thanks to the eXn group members at the University of Houston. 

\end{acknowledgments}

\bibliographystyle{unsrt}
\bibliography{main}% Produces the bibliography via BibTeX.

\providecommand{\noopsort}[1]{}\providecommand{\singleletter}[1]{#1}%
\begin{thebibliography}{10}

\bibitem{mondal2020magnetic}
Suchanda Mondal, A~Midya, Manju~Mishra Patidar, V~Ganesan, and Prabhat Mandal.
\newblock Magnetic and magnetocaloric properties of layered van der waals
  {CrCl$_3$}.
\newblock {\em Applied Physics Letters}, 117(9), 2020.

\bibitem{tsubokawa1960magnetic}
Ichiro Tsubokawa.
\newblock On the magnetic properties of a {CrBr$_3$} single crystal.
\newblock {\em Journal of the Physical Society of Japan}, 15(9):1664--1668,
  1960.

\bibitem{mcguire2015coupling}
Michael~A McGuire, Hemant Dixit, Valentino~R Cooper, and Brian~C Sales.
\newblock Coupling of crystal structure and magnetism in the layered,
  ferromagnetic insulator {CrI$_3$}.
\newblock {\em Chemistry of Materials}, 27(2):612--620, 2015.

\bibitem{shao2021spectroscopic}
YC~Shao, B~Karki, W~Huang, X~Feng, G~Sumanasekera, J-H Guo, Y-D Chuang, and
  B~Freelon.
\newblock Spectroscopic determination of key energy scales for the base
  hamiltonian of chromium trihalides.
\newblock {\em The Journal of Physical Chemistry Letters}, 12(1):724--731,
  2021.

\bibitem{craco2021electronic}
L~Craco, SS~Carara, Y-C Shao, Y-D Chuang, and B~Freelon.
\newblock Electronic structure of rhombohedral {CrX$_3$ (X= Br, Cl, I)} van der
  waals crystals.
\newblock {\em Physical Review B}, 103(23):235119, 2021.

\bibitem{boeyens2008periodic}
Jan~CA Boeyens.
\newblock The periodic electronegativity table.
\newblock {\em Zeitschrift f{\"u}r Naturforschung B}, 63(2):199--209, 2008.

\bibitem{goodenough1963magnetism}
John~B Goodenough.
\newblock {\em Magnetism and the chemical bond}.
\newblock R. E. Krieger Pub. Co., 1963.

\bibitem{mermin1966absence}
N~David Mermin and Herbert Wagner.
\newblock Absence of ferromagnetism or antiferromagnetism in one-or
  two-dimensional isotropic heisenberg models.
\newblock {\em Physical Review Letters}, 17(22):1133, 1966.

\bibitem{ghosh2023magnetic}
Anirudha Ghosh, H~Johan~M J{\"o}nsson, Deepak~John Mukkattukavil, Yaroslav
  Kvashnin, Dibya Phuyal, Patrik Thunstr{\"o}m, Marcus Ag{\aa}ker, Alessandro
  Nicolaou, Martin Jonak, R{\"u}diger Klingeler, et~al.
\newblock Magnetic circular dichroism in the $dd$ excitation in the van der
  waals magnet {CrI$_3$} probed by resonant inelastic {X}-ray scattering.
\newblock {\em Physical Review B}, 107(11):115148, 2023.

\bibitem{atanasov2012modern}
Mihail Atanasov, Dmitry Ganyushin, Kantharuban Sivalingam, and Frank Neese.
\newblock A modern first-principles view on ligand field theory through the
  eyes of correlated multireference wavefunctions.
\newblock {\em Molecular Electronic Structures of Transition Metal Complexes
  II}, pages 149--220, 2012.

\bibitem{jiang2018controlling}
Shengwei Jiang, Lizhong Li, Zefang Wang, Kin~Fai Mak, and Jie Shan.
\newblock Controlling magnetism in 2d {CrI$_3$} by electrostatic doping.
\newblock {\em Nature Nanotechnology}, 13(7):549--553, 2018.

\bibitem{novoselov20162d}
K~S Novoselov, Artem Mishchenko, Alexandra Carvalho, and AH~Castro~Neto.
\newblock 2d materials and van der waals heterostructures.
\newblock {\em Science}, 353(6298):aac9439, 2016.

\bibitem{pan2018large}
Longfei Pan, Le~Huang, Mianzeng Zhong, Xiang-Wei Jiang, Hui-Xiong Deng, Jingbo
  Li, Jian-Bai Xia, and Zhongming Wei.
\newblock Large tunneling magnetoresistance in magnetic tunneling junctions
  based on two-dimensional {CrX$_3$ (X= Br, I)} monolayers.
\newblock {\em Nanoscale}, 10(47):22196--22202, 2018.

\bibitem{akram2021moire}
Muhammad Akram, Harrison LaBollita, Dibyendu Dey, Jesse Kapeghian, Onur Erten,
  and Antia~S Botana.
\newblock Moir{\'e} skyrmions and chiral magnetic phases in twisted {CrX$_3$
  (X= I, Br, and Cl)} bilayers.
\newblock {\em Nano Letters}, 21(15):6633--6639, 2021.

\bibitem{singamaneni2020light}
SR~Singamaneni, LM~Martinez, J~Niklas, OG~Poluektov, R~Yadav, M~Pizzochero,
  OV~Yazyev, and MA~McGuire.
\newblock Light induced electron spin resonance properties of van der waals
  {CrX$_3$ (X= Cl, I)} crystals.
\newblock {\em Applied Physics Letters}, 117(8), 2020.

\bibitem{mishra2024magnetic}
Prakash Mishra and Tunna Baruah.
\newblock Magnetic properties of {CrX$_3$ (X= Cl, Br, I)} monolayers in excited
  states.
\newblock {\em Journal of Materials Chemistry C}, 12:5216--5221, 2024.

\bibitem{ahmad2020pressure}
Azkar~Saeed Ahmad, Yongcheng Liang, Mingdong Dong, Xuefeng Zhou, Leiming Fang,
  Yuanhua Xia, Jianhong Dai, Xiaozhi Yan, Xiaohui Yu, Junfeng Dai, et~al.
\newblock Pressure-driven switching of magnetism in layered {CrCl$_3$}.
\newblock {\em Nanoscale}, 12(45):22935--22944, 2020.

\bibitem{pollini1970intrinsic}
I~Pollini and G~Spinolo.
\newblock Intrinsic optical properties of {CrCl$_3$}.
\newblock {\em Physica status solidi (b)}, 41(2):691--701, 1970.

\bibitem{dillon1963magneto}
JF~Dillon~Jr, H~Kamimura, and JP~Remeika.
\newblock Magneto-optical studies of chromium tribromide.
\newblock {\em Journal of Applied Physics}, 34(4):1240--1245, 1963.

\bibitem{limido1972specific}
C~Limido, G~Pedroli, and G~Spinolo.
\newblock Specific magnetic rotation in the crystal field bands of {CrCl$_3$}.
\newblock {\em Solid State Communications}, 11(10):1385--1388, 1972.

\bibitem{bermudez1979spectroscopic}
Victor~M Bermudez and Donald~S McClure.
\newblock Spectroscopic studies of the two-dimensional magnetic insulators
  chromium trichloride and chromium tribromide—{I}.
\newblock {\em Journal of Physics and Chemistry of Solids}, 40(2):129--147,
  1979.

\bibitem{suganotanabe}
Satoru Sugano.
\newblock {\em Multiplets of Transition-Metal Ions in Crystals}.
\newblock Elsevier, 1970.

\bibitem{katsnelson2010theory}
MI~Katsnelson and AI~Lichtenstein.
\newblock Theory of optically forbidden $d-d$ transitions in strongly
  correlated crystals.
\newblock {\em Journal of Physics: Condensed Matter}, 22(38):382201, 2010.

\bibitem{Singh:yi5098}
A.~Singh, H.~Y. Huang, Y.~Y. Chu, C.~Y. Hua, S.~W. Lin, H.~S. Fung, H.~W. Shiu,
  J.~Chang, J.~H. Li, J.~Okamoto, C.~C. Chiu, C.~H. Chang, W.~B. Wu, S.~Y.
  Perng, S.~C. Chung, K.~Y. Kao, S.~C. Yeh, H.~Y. Chao, J.~H. Chen, D.~J.
  Huang, and C.~T. Chen.
\newblock {Development of the Soft {X}-ray AGM{-}AGS RIXS beamline at the
  Taiwan Photon Source}.
\newblock {\em Journal of Synchrotron Radiation}, 28(3):977--986, May 2021.

\bibitem{ament2011resonant}
Luuk~JP Ament, Michel Van~Veenendaal, Thomas~P Devereaux, John~P Hill, and
  Jeroen Van Den~Brink.
\newblock Resonant inelastic {X}-ray scattering studies of elementary
  excitations.
\newblock {\em Reviews of Modern Physics}, 83(2):705, 2011.

\bibitem{occhialini2025spin}
Connor~A Occhialini, Luca Nessi, Luiz~GP Martins, Ahmet~Kemal Demir, Qian Song,
  Vicky Hasse, Chandra Shekhar, Claudia Felser, Kenji Watanabe, Takashi
  Taniguchi, et~al.
\newblock Spin-forbidden excitations in the magneto-optical spectra of
  {CrI$_3$} tuned by covalency.
\newblock {\em Physical Review X}, 15(3):031005, 2025.

\bibitem{BAKER2017182}
Michael~L. Baker, Michael~W. Mara, James~J. Yan, Keith~O. Hodgson, Britt
  Hedman, and Edward~I. Solomon.
\newblock K- and l-edge {X}-ray absorption spectroscopy (xas) and resonant
  inelastic {X}-ray scattering (rixs) determination of differential orbital
  covalency (doc) of transition metal sites.
\newblock {\em Coordination Chemistry Reviews}, 345:182--208, 2017.
\newblock Chemical Bonding: "State of the Art".

\bibitem{wang2017charge}
Ru-Pan Wang, Boyang Liu, Robert~J Green, Mario~Ulises Delgado-Jaime, Mahnaz
  Ghiasi, Thorsten Schmitt, Matti~M van Schooneveld, and Frank~MF De~Groot.
\newblock Charge-transfer analysis of $2p3d$ resonant inelastic {X}-ray
  scattering of cobalt sulfide and halides.
\newblock {\em The Journal of Physical Chemistry C}, 121(45):24919--24928,
  2017.

\bibitem{haverkort2016quanty}
Maurits~W Haverkort.
\newblock Quanty for core level spectroscopy-excitons, resonances and band
  excitations in time and frequency domain.
\newblock In {\em Journal of Physics: Conference Series}, volume 712, page
  012001. IOP Publishing, 2016.

\bibitem{crispy}
Marius Retegan and Stephan Kuschel.
\newblock mretegan/crispy: v0.7.4, 2023.

\bibitem{haverkort2012multiplet}
MW~Haverkort, M~Zwierzycki, and OK~Andersen.
\newblock Multiplet ligand-field theory using wannier orbitals.
\newblock {\em Physical Review B}, 85(16):165113, 2012.

\bibitem{de2008core}
Frank De~Groot and Akio Kotani.
\newblock {\em Core level spectroscopy of solids}.
\newblock CRC press, 2008.

\bibitem{hunault2018direct}
Myrtille~OJY Hunault, Yoshihisa Harada, Jun Miyawaki, Jian Wang, Andries
  Meijerink, Frank~MF De~Groot, and Matti~M Van~Schooneveld.
\newblock Direct observation of {Cr$^{3+}$} $3d$ states in ruby: Toward
  experimental mechanistic evidence of metal chemistry.
\newblock {\em The Journal of Physical Chemistry A}, 122(18):4399--4413, 2018.

\bibitem{racah1942theory}
Giulio Racah.
\newblock Theory of complex spectra. {I}.
\newblock {\em Physical Review}, 61(3-4):186, 1942.

\bibitem{hunault2016effect}
Myrtille~OJY Hunault, Laurence Galoisy, G{\'e}rald Lelong, Matt Newville, and
  Georges Calas.
\newblock Effect of cation field strength on {Co$^{2+}$} speciation in
  alkali-borate glasses.
\newblock {\em Journal of Non-Crystalline Solids}, 451:101--110, 2016.

\bibitem{seyler2018ligand}
Kyle~L Seyler, Ding Zhong, Dahlia~R Klein, Shiyuan Gao, Xiaoou Zhang, Bevin
  Huang, Efr{\'e}n Navarro-Moratalla, Li~Yang, David~H Cobden, Michael~A
  McGuire, et~al.
\newblock Ligand-field helical luminescence in a 2d ferromagnetic insulator.
\newblock {\em Nature Physics}, 14(3):277--281, 2018.

\bibitem{he2025dispersive}
W~He, J~Sears, F~Barantani, T~Kim, JW~Villanova, T~Berlijn, M~Lajer,
  MA~McGuire, J~Pelliciari, V~Bisogni, et~al.
\newblock Dispersive dark excitons in van der waals ferromagnet {CrI$_3$}.
\newblock {\em Physical Review X}, 15(1):011042, 2025.

\bibitem{DILLON19661531}
J.F. Dillon, H.~Kamimura, and J.P. Remeika.
\newblock Magneto-optical properties of ferromagnetic chromium trihalides.
\newblock {\em Journal of Physics and Chemistry of Solids}, 27(9):1531--1549,
  1966.

\bibitem{juhin2008x}
Am{\'e}lie Juhin, Christian Brouder, Marie-Anne Arrio, Delphine Cabaret,
  Philippe Sainctavit, Etienne Balan, Am{\'e}lie Bordage, Ari~P Seitsonen,
  Georges Calas, Sigrid~G Eeckhout, et~al.
\newblock {X}-ray linear dichroism in cubic compounds: The case of {Cr$^{3+}$}
  in {MgAl$_2$O$_4$}.
\newblock {\em Physical Review B}, 78(19):195103, 2008.

\bibitem{serri2020enhancement}
Michele Serri, Giuseppe Cucinotta, Lorenzo Poggini, Giulia Serrano, Philippe
  Sainctavit, Judyta Strychalska-Nowak, Antonio Politano, Francesco Bonaccorso,
  Andrea Caneschi, Robert~J Cava, et~al.
\newblock Enhancement of the magnetic coupling in exfoliated {CrCl$_3$}
  crystals observed by low-temperature magnetic force microscopy and {X}-ray
  magnetic circular dichroism.
\newblock {\em Advanced Materials}, 32(24):2000566, 2020.

\bibitem{vercamer2016calculation}
Vincent Vercamer, Myrtille~OJY Hunault, G{\'e}rald Lelong, Maurits~W Haverkort,
  Georges Calas, Yusuke Arai, Hiroyuki Hijiya, Lorenzo Paulatto, Christian
  Brouder, Marie-Anne Arrio, et~al.
\newblock Calculation of optical and {K} pre-edge absorption spectra for
  ferrous iron of distorted sites in oxide crystals.
\newblock {\em Physical Review B}, 94(24):245115, 2016.

\bibitem{feldkemper1998generalized}
Stefan Feldkemper and Werner Weber.
\newblock Generalized calculation of magnetic coupling constants for
  mott-hubbard insulators: Application to ferromagnetic {Cr} compounds.
\newblock {\em Physical Review B}, 57(13):7755, 1998.

\bibitem{dalal2017textbook}
Mandeep Dalal.
\newblock {\em A Textbook of Inorganic Chemistry - Volume 1}.
\newblock Dalal Institute, 2017.

\bibitem{schlafer1969basic}
Hans~Ludwig Schl{\"a}fer and G{\"u}nter Gliemann.
\newblock {\em Basic principles of ligand field theory}.
\newblock Wiley-Interscience, 1969.

\bibitem{pollini1989optical}
I~Pollini, J~Thomas, B~Carricaburu, and R~Mamy.
\newblock Optical and electron energy loss experiments in ionic {CrBr$_3$}
  crystals.
\newblock {\em Journal of Physics: Condensed Matter}, 1(41):7695, 1989.

\bibitem{acharya2021excitons}
Swagata Acharya, Dimitar Pashov, Alexander~N Rudenko, Malte R{\"o}sner, Mark
  van Schilfgaarde, and Mikhail~I Katsnelson.
\newblock Excitons in bulk and layered chromium tri-halides: from frenkel to
  the wannier-mott limit.
\newblock {\em arXiv preprint arXiv:2110.08174}, 2021.

\bibitem{van19922p}
G~Van~der Laan and IW~Kirkman.
\newblock The $2p$ absorption spectra of $3d$ transition metal compounds in
  tetrahedral and octahedral symmetry.
\newblock {\em Journal of Physics: Condensed Matter}, 4(16):4189, 1992.

\bibitem{zheng2011theoretical}
Wen-Chen Zheng, Gu-Ming Jia, Lv~He, and Wei-Qing Yang.
\newblock A theoretical study on the temperature dependence of zero-field
  splitting for the tetragonal {Cr$^{3+}$} center in {MgO} crystal.
\newblock {\em Spectrochimica Acta Part A: Molecular and Biomolecular
  Spectroscopy}, 78(2):818--820, 2011.

\bibitem{kvashnin2020relativistic}
YO~Kvashnin, Anders Bergman, AI~Lichtenstein, and MI~Katsnelson.
\newblock Relativistic exchange interactions in {Cr$X_3$ ($X$ = Cl, Br, I)}
  monolayers.
\newblock {\em Physical Review B}, 102(11):115162, 2020.

\bibitem{molina2020magneto}
Alejandro Molina-S{\'a}nchez, Gon{\c{c}}alo Catarina, Davide Sangalli, and
  Joaquin Fernandez-Rossier.
\newblock Magneto-optical response of chromium trihalide monolayers: chemical
  trends.
\newblock {\em Journal of Materials Chemistry C}, 8(26):8856--8863, 2020.

\bibitem{dillon1965magnetization}
JF~Dillon~Jr and CE~Olson.
\newblock Magnetization, resonance, and optical properties of the ferromagnet
  {CrI$_3$}.
\newblock {\em Journal of Applied Physics}, 36(3):1259--1260, 1965.

\bibitem{liu2023light}
Xiaopeng Liu, Dominik Legut, and Qianfan Zhang.
\newblock Light-induced ultrafast enhancement of magnetic orders in monolayer
  {CrX$_3$}.
\newblock {\em The Journal of Physical Chemistry C}, 127(27):13398--13406,
  2023.

\bibitem{tchougreeff2009nephelauxetic}
Andrei~L Tchougr{\'e}eff and Richard Dronskowski.
\newblock Nephelauxetic effect revisited.
\newblock {\em International Journal of Quantum Chemistry}, 109(11):2606--2621,
  2009.

\bibitem{jimenez2015x}
J~Jim{\'e}nez-Mier, P~Olalde-Velasco, W-L Yang, and J~Denlinger.
\newblock X-ray absorption and resonant inelastic x-ray scattering (rixs) show
  the presence of cr+ at the surface and in the bulk of crf2.
\newblock In {\em AIP Conference Proceedings}, volume 1671, page 020002. AIP
  Publishing LLC, 2015.

\bibitem{sant2023anisotropic}
Roberto Sant, Alessandro De~Vita, Vincent Polewczyk, Gian~Marco Pierantozzi,
  Federico Mazzola, Giovanni Vinai, Gerrit van~der Laan, Giancarlo Panaccione,
  and NB~Brookes.
\newblock Anisotropic hybridization probed by polarization dependent {X}-ray
  absorption spectroscopy in {VI$_3$} van der waals mott ferromagnet.
\newblock {\em Journal of Physics: Condensed Matter}, 35(40):405601, 2023.

\bibitem{camerano2024symmetry}
Luigi Camerano and Gianni Profeta.
\newblock Symmetry breaking in vanadium trihalides.
\newblock {\em 2D Materials}, 11(2):025027, 2024.

\bibitem{tomiyasu2017coulomb}
Keisuke Tomiyasu, Jun Okamoto, Hsiao-Yu Huang, Zhi-Ying Chen, Evelyn~Pratami
  Sinaga, Wen-Bin Wu, Yen-Yi Chu, Amol Singh, R-P Wang, FMF De~Groot, et~al.
\newblock Coulomb correlations intertwined with spin and orbital excitations in
  {LaCoO$_3$}.
\newblock {\em Physical Review Letters}, 119(19):196402, 2017.

\bibitem{miyawaki2017dzyaloshinskii}
Jun Miyawaki, Shigemasa Suga, Hidenori Fujiwara, Masato Urasaki, Hidekazu
  Ikeno, Hideharu Niwa, Hisao Kiuchi, and Yoshihisa Harada.
\newblock Dzyaloshinskii-moriya interaction in $\alpha$-{$Fe_2O_3$} measured by
  magnetic circular dichroism in resonant inelastic soft {X}-ray scattering.
\newblock {\em Physical Review B}, 96(21):214420, 2017.

\end{thebibliography}
\end{document}